\documentclass[11pt]{article}

\usepackage{amsmath, amssymb, wasysym, slashed, multirow,hyperref}
\usepackage{pdfpages}
\usepackage{cite}
\usepackage{times}

\newenvironment{sciabstract}{%
\begin{quote} }
{\end{quote}}

\newcounter{lastnote}

\title{Resummation and renormalons in a general Quantum Field Theory}

\author
{Alessio Maiezza,$^{1\ast}$  Juan Carlos Vasquez$^{2\dagger}$\\
\\
\normalsize{$^{1}$Ruder Bo\v skovi\'c Institute, Bijeni\v cka cesta 54, 10000, Zagreb, Croatia,}\\
\normalsize{$^{2}$Universidad T\'ecnica Federico Santa Mar\'ia $\&$ CCTVal, Valpara\'iso, Chile}\\
\\
\small{ E-mail: amaiezza@irb.hr$^{\ast}$,juan.vasquezcar@usm.cl$^{\dagger}$}
}

\date{}

\begin{document}


\baselineskip16pt 


\maketitle


\begin{sciabstract}
                  We generalize the concept of Borel resummability and renormalons to a quantum field theory with an arbitrary number of fields and couplings, starting from the known notion based on the running coupling constants. An approach to identify the renormalons is provided by exploiting an analytic solution of the generic one-loop renormalization group equations in multi-field theories. Methods to evaluate the regions in coupling space where the theory is resummable are described. The generalization is then illustrated in a toy model with two coupled scalar fields, representing the simplest extension of the one-field analysis presented in the seminal works of the subject. Furthermore, possible links to realistic theories are briefly discussed.
\end{sciabstract}


\section{Introduction}

It has been known since the original argument given by Dyson~\cite{Dyson:1952tj}, that the all the power series used in Quantum Field Theory (QFT) diverge as $n!$, where $n$ is the order in the perturbative expansion.
However,  the convergence of a divergent series can be improved through the Borel transform, followed by an analytic continuation  and finally coming back to the original variable and a finite output through the Laplace transform.
This procedure is called resummation and helps to make sense of the divergent series in QFT. Unfortunately,  even the Borel series may diverge in some cases~\cite{tHooft:1977xjm, Parisi:1978bj,Parisi:1978iq,Lipatov}. This divergence  may be due  to the instantons~\cite{Lipatov}, but these may be treated consistently by semiclassical methods~\cite{tHooft:1977xjm,Coleman:1977py,Bogomolny:1980ur,ZinnJustin:1981dx}.

The real problem is due to another type of divergence of the Borel series, the so-called renormalons~\cite{tHooft:1977xjm},  whose physics is not well understood.
As suggested by the name first introduced by 't Hooft, the renormalons arise from the procedure of renormalization and bring in ambiguities in the perturbative formulation of the theory when the coupling is large enough (for complete reviews see~\cite{LeGuillou:1990nq,Beneke:1998ui,Weinberg_book,Shifman:2013uka}).
Since a rigorous proof of the existence of renormalons has not yet been found, there is a claim that they may not even exist at all~\cite{Suslov:2005zi}. Nevertheless, in Ref.~\cite{David:1988bi} the existence of renormalons was established  for the $N$ component $\phi^4$ scalar field theory. Moreover,  in Ref.~\cite{Bauer:2011ws}  compelling evidence of the existence of the factorial growth predicted by renormalon analysis has also been found in QCD, where  the renormalons have been used to estimate the non-perturbative power corrections.~\cite{Beneke:1994sw,Bigi:1994em}.

In this work, we propose the generalization of the original concept of renormalons to a multi-field and multi-coupling framework starting from the argument presented in~\cite{tHooft:1977xjm} and connected with the renormalization group in~\cite{Parisi:1978bj,Parisi:1978iq}, the latter offering a natural way to extend it through the notion of the running couplings. The enlargement to multi-field theories in non-trivial in several aspects. A relevant obstacle is to obtain the analytic solution of renormalization-group-equations~\cite{GellMann:1954fq} (RGEs). A coupled system of differential equations is in general not solvable analytically. We circumvent the problem by exploiting an iterative solution of the RGEs. This enables us to define a multi-variable Borel and Laplace transforms, to estimate the singularity of the former and to identify the renormalons related to the ambiguity of the latter. Particular attention is paid to scalar field theories. In order to illustrate the proposed method, we study a simple prototype model of two coupled scalar fields, the minimal generalization of the example in Ref.~\cite{tHooft:1977xjm}.

The article is organized as follows: in Sec.~\ref{onefield}, for the reader's ease, we review the main features of the singularities of the Borel transform with a particular focus on renormalons. In Sec.~\ref{RGEs} we show the recursive and analytical solution for the RGEs. Then in Sec.~\ref{multi-field-theory} we construct the general  Borel transform and show how to estimate its singularities, within different approximations, analytically or numerically. In Sec.~\ref{toy_model} and for the sake of illustration, we apply the method previously discussed for a scalar toy model. In Sec.~\ref{contact}, we comment on the contact with realistic models and conclusions are given in Sec.~\ref{outlook}. Finally, the paper is equipped with several Appendices for further details.

\section{Renormalons: main features }\label{onefield}

In this section we review the main features of the singularities of the Borel transform, known as instantons and renormalons, disentangling them and focusing on the latter.
The aim is to define a starting point and a road-map for the generalization of renormalons in any QFT through the Borel resummability in multi-variables series.

\subsection{Divergences of Borel serie: instantons vs renormalons}

The procedure of resummation improves the convergence of the perturbation series in QFT. This is achieved through the Borel transform
of those series (see Ref.~\cite{Resurgence} for the mathematical theory of resurgent analysis). To be more explicit and to provide a set-up, it may be useful to start considering the over-simplistic case of a QFT in one space-time point~\cite{tHooft:1977xjm}, so that the functional integral reduces  to
\begin{equation}
G(\lambda)= \int dx e^{-\frac{1}{2}x^2-\frac{\lambda}{4!}x^4}\,.
\end{equation}
This function can be re-written as a Laplace transform
\begin{equation}
G(\lambda)=\int_0^\infty dz F(z)e^{-z /\lambda}\,,
\end{equation}
where $F(z)$ is the Borel transform. Thus the Laplace transform is well-defined if  $F(z)$ has no poles on the real and positive axes, otherwise, the usual perturbative approach to the theory becomes ambiguous. The ambiguity is due to the two possible contour deformations that one may choose in order to avoid the pole in the positive real axes when applying the inverse Borel transform.
An insight regarding  the position of the pole in  $F(z)$ can be obtained by  writing  it  as
\begin{equation}
F(z) = \int dx \delta(z-S(x)/\lambda)\,,
 \end{equation}
with the $S(x)=-\frac{1}{2}x^2-\frac{\lambda}{4!}x^4$.
Notice that  it diverges when $\sim dS/dx|_{x=\bar{x}}=0$, where $\bar{x}$ satisfies $S(\bar{x})=\lambda z$~\cite{tHooft:1977xjm} (see~\cite{Flory:2012nk} for a  discussion in one dimension, and~\cite{Serone:2016qog,Serone:2017nmd} for the analogy with the anharmonic oscillator in quantum mechanics). Hence,  in order to locate the divergence of the Borel transform $F(z)$ one must look for a solution of the classical equations of motion.  Such solution is known as instanton and it is related to barrier penetration~\cite{Vainshtein:1964dw,Coleman:1977py}. Although being a non-perturbative object, is not retained dangerous since it can be treated semi-classically~\cite{Coleman:1977py,Bogomolny:1980ur,ZinnJustin:1981dx}. The discussion here is based on the one-dimensional integral and holds in the multidimensional case~\cite{tHooft:1977xjm}. Moreover, the position of the singularity is not affected by the usual manipulations in QFT, such  as ratios or multiplication of such integrals and differentiation with respect to the source~\cite{tHooft:1977xjm}.

In a renormalizable QFT, there is another class of contributions to the Green's functions growing faster than $n!$. These contributions arise after  reabsorbing  the infinite terms at the $n$th-order  perturbative calculation and hence have no analogy with the one-dimensional example sketched before. In other words, such contributions are a byproduct of the renormalization procedure and  for this reason  are called renormalons. These are briefly reviewed in the next subsection and then generalized to several fields and couplings.

 \begin{figure}\centering
     \includegraphics[scale=0.5]{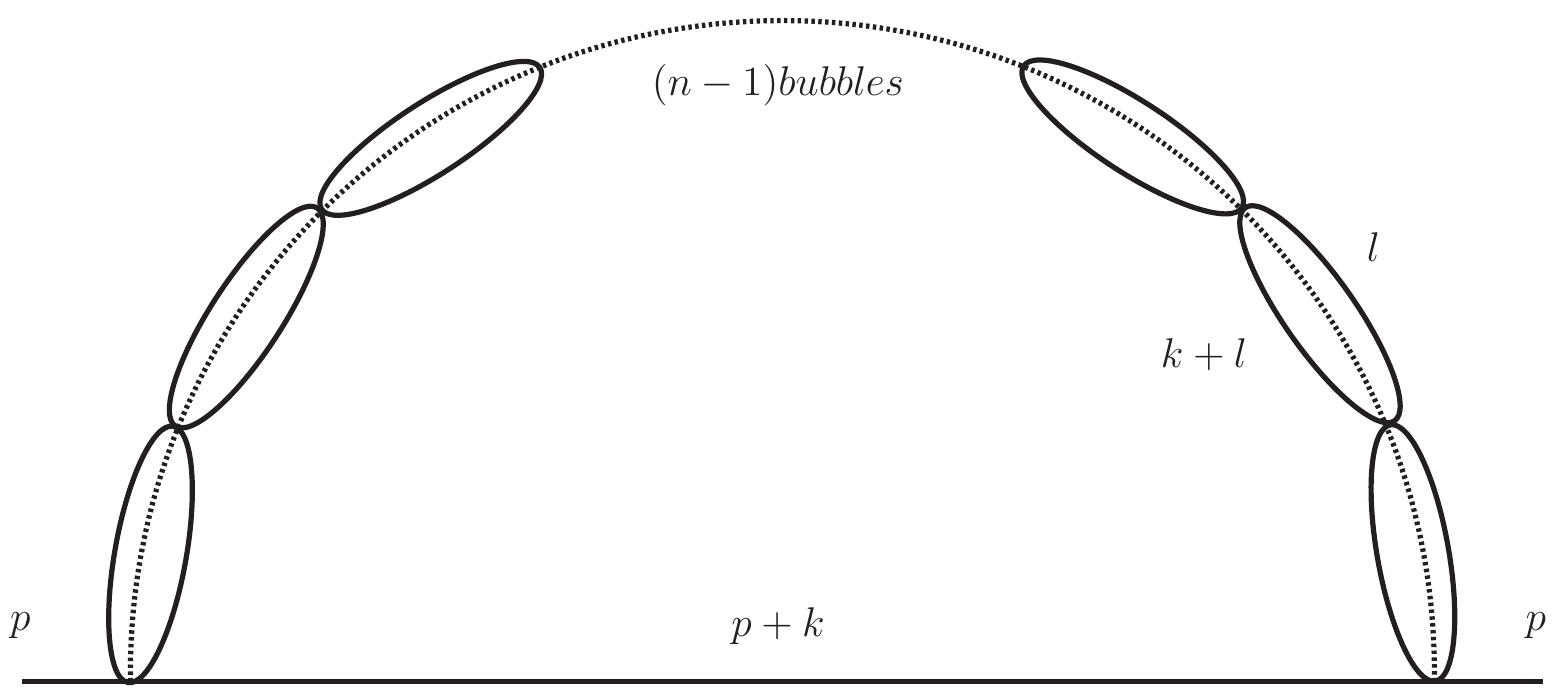}
      \includegraphics[scale=0.4]{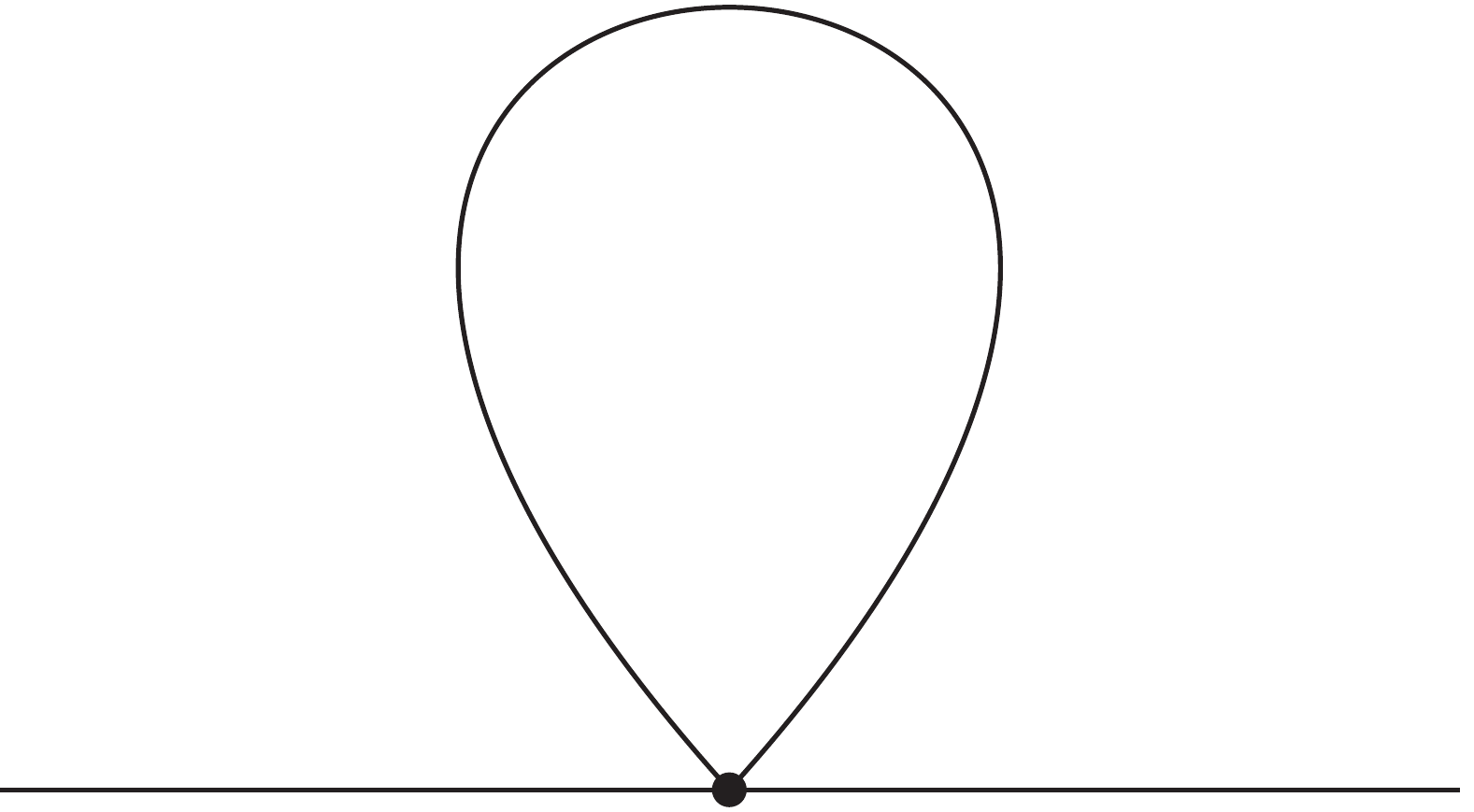}
  \caption{Alternative understanding of the n! growth renormalon contribution.
  Left. Feynman  diagram for the renormalon in $\lambda\phi^4$ theory. The dashed line represents the ''one-loop skeleton''
  with (n-1) loops. The  lines  represent  the scalars, to distinguish them  from the skeleton structure. Right. One loop correction to the quartic vertex leading to renormalon singularity in scalar field theory. The black dot represents the one-loop running coupling in the vertex.}
  \label{fig:Renormalon}
\end{figure}

\subsection{Renormalons in the one field case}

In this section we review the main features of the renormalon divergences in the one-field case. The original argument~\cite{tHooft:1977xjm} is based on the
Feynman  diagram on Fig.~\ref{fig:Renormalon} (left)\footnote{The same diagram has been considered in~\cite{Loewe:1999kw} to include thermal corrections to renormalons  in scalar field theories.} for   $\frac{\lambda}{4!}\phi^4$ theory, given in term of the one-loop bubble function $B(k)$:
\begin{align}\label{Rn}
&R_n = \int \frac{d^4k}{(2\pi)^4}\frac{i}{(p+k)^2-m^2+i\epsilon}\frac{1}{(-i\lambda)^{n-1}}[B(k)]^{n}\,,\\
&B(k)= \frac{(-i\lambda)^2}{2}\int\frac{d^4l}{(2\pi)^4}\frac{i}{(k+l)^2-m^2+i\epsilon}\frac{i}{l^2-m^2+i\epsilon}\,. \nonumber
\end{align}
Expanding $B(k)$ for large Euclidean momentum $k_E\gg m$ and after reabsorbing the infinite terms into the mass and the quartic coupling, the expression at leading order in $\mathcal{O}(p/m)$ for $R_n$ is~\cite{tHooft:1977xjm}
\begin{equation}\label{factors}
R_n \sim i\lambda^{n+1}\left(\frac{\beta}{2}\right)^{n}\frac{1}{16}\frac{m^4}{\mu^2}n! \,
\end{equation}
where $\beta$ is defined by the one loop RGE $\lambda'=\beta\lambda^2$ and  the prime denotes the  derivative with respect to $t=\ln(\mu/\mu_0)$.

The formal Borel transform $\mathcal{B}$ of the term $\lambda^{n+1}$ is given by  $\mathcal{B}^{(1)}(\lambda^{n+1}) \equiv \frac{z^n}{n!} $
where the index on $\mathcal{B}$ refers to one-field (coupling).
Applying the Borel transform to Eq.~\eqref{running_Renormalon} one finds
\begin{equation}
\mathcal{B}^{(1)}(\sum_n R_n)\propto \sum_n(\frac{\beta}{2}z)^n \equiv \sum_n \mathcal{B}^{(1)}_n(z) =\frac{1}{1-\frac{\beta}{2}z}\,, \label{Renormalon_1field}
\end{equation}
and the Borel series diverges  at
\begin{equation}\label{onefieldRenormalon}
z_{pole}=2/\beta\,.
\end{equation}
Such divergence is on the real and positive axis for $\beta>0$ and it represents the so called ultraviolet renormalon. Moreover, there are an infinite number of these singularities discretized in units of $2/\beta$~\cite{tHooft:1977xjm,Parisi:1978bj,Parisi:1978iq}, thus larger than the one in Eq.~\eqref{onefieldRenormalon}.
The perturbative expansion can be consistently used for values of the coupling $\lambda\ll \frac{2}{\beta}$. There is another kind of renormalon singularity when the momentum $k\rightarrow 0$ that are called infrared renormalons that we do not further discuss.

On a more general ground, one can understand such divergences in terms of renormalization theory~\cite{Parisi:1978bj,Parisi:1978iq}. In the specific case illustrated here,
the $n!$ growth and the divergence~\eqref{onefieldRenormalon} of the diagram in Fig.~\ref{fig:Renormalon} (left) is equivalent to the one in the one-loop diagram in Fig.~\ref{fig:Renormalon} (right) with the running coupling in the vertex (see for example~\cite{Beneke:1998ui})
\begin{align}
\sum_n R_n &=\sum_ni\lambda^{n+1}\left(\frac{\beta}{2}\right)^{n}\frac{1}{16}\frac{m^4}{\mu_0^2}n!  = \frac{im^4}{8\pi^2}\int \frac{d k_E^2}{2(k_E^2)^2}\frac{\lambda(\mu_0)}{1-\frac{\beta}{2} \lambda(\mu_0)^2\ln(k_E^2/\mu_0^2)}  \nonumber \\
& \simeq \frac{im^4}{8\pi^2}\int \frac{d k_E^2}{2(k_E^2)^2}\lambda(k_E) \,.\label{running_Renormalon}
\end{align}
The last expression can be thought as an effective-RG-improved vertex~\cite{Coleman:1973jx,Ford:1992mv} and thus is general. Therefore the RGEs represent the heart of the approach to the extension of Borel resummability and renormalons in multi-variables. After having constructed an recursive-analytical solution of multi-coupling RGEs, we shall explicitly show that the $n!$ growth still holds.

\section{Analytical one-loop RGEs in multi-field theory} \label{RGEs}

In multi-field theories, the RGEs solutions tangle drastically and are often dealt with a numerical approach.
As discussed above, one needs  an analytical form of RGEs solution in the multi-coupling case,
that to our knowledge has never been tackled in the literature.

Let us first consider scalar fields and write the RGEs in the generic form
\begin{equation}\label{}
\lambda_i'=\sum_{n,m=1}^N \beta_i^{nm} \lambda_n \lambda_m \equiv \beta_i^{nm} \lambda_n \lambda_m\,,
\end{equation}
the prime denotes derivative with respect to $t=\ln(\mu/\mu_0)$ and $N$ is  the number of couplings. We adopt the summation convention over repeated indices and when confusion arises we explicitly denote the  summation. All the latin indices $i,j,...$ run from 1 to $N$ and for convenience we generically  denote the $\beta_i^{nm}$ coefficients as $\beta$s.

The  solution may be written recursively from the generalized variational method~\cite{he2000variational} as
\begin{equation}\label{rec_int}
\lambda_i^{(n)}(t) = \lambda_i^{(n-1)}(t) - \int_0^t \left( \frac{d\lambda_i^{(n-1)}}{ds} -  \sum_{j,k}^N \beta_i^{j k} \lambda_j^{(n-1)}(s) \lambda_k^{(n-1)}(s)  \right)ds\,.
\end{equation}
The exact solution is formally obtained when $n\rightarrow \infty$ and  $\lambda_i^{(0)}(t)= \lambda_i^{(0)}(0)$, i.e. it is the initial condition. This is the complete analytical solution of the RGEs at one-loop. It provides the required expansion in the variable $t$ and is the equivalent of the one-field case but approximated, in the sense that in practice one stops at a given order in powers of $t$. Clearly, in the generic case the coefficient of the term $t^n$ is more complicated than the one-field case, but the $n!$ divergence encountered in section~\ref{onefield} is left intact.

The expression shown in Eq.~\eqref{rec_int} can be formally integrated and the
solution at a given order $n$ can be  arranged as
\begin{align}\label{solution}
\lambda_i(t)&= \lambda_i(0)+ tv_i+M_i^k s(1)_kt^2+ \sum_{n=3}^{\infty}s(n)_i \frac{t^n}{n}\nonumber \\
s(n)_i&=  M_i^k  s(n-1)_k+\beta_i^{kl} \sum_{m=1}^{n-2}s(m)_ks(n-1-m)_l\,,
\end{align}
with $s(1)_i \equiv v_i$ and where we have defined
\begin{equation}
v_i \equiv \beta_i^{mn}  \lambda_m(0) \lambda_n(0) , \quad M_i^k=\frac{\partial s(1)_i }{\partial \lambda_k}\equiv( \beta_i^{mk}+\beta_i^{km}) \lambda_m(0)\,.\quad
\end{equation}
As usual, the couplings within this operators are defined at the scale $t=0$ or $\mu=\mu_0$. The iterative solution in the form of Eq.~\eqref{solution} is ready for a quick implementation up to any order $n$.

For the reader's sake, in Appendix~\ref{SecApp1} and for a simple toy model, we compare the recursive analytical solution in Eq.~\eqref{rec_int} with the usual numeric one. The treatment applied to Eq.~\eqref{rec_int} can be generalized to equations including cubic and quartic terms, necessary to take into account fermion fields in the RGEs. This is explicitly done in Appendix~\ref{generalRGE}.

\section{Renormalons in multi-field theory} \label{multi-field-theory}

As discussed in Sec.~\ref{onefield}, the analytical solution of RGEs and in particular the expression~\eqref{running_Renormalon} supplies a straightforward insight for a generalization to multi-variable of renormalons singularity. From the diagram in Fig.~\ref{fig:Renormalon} (right) one has
\begin{align}
 \text{Eq.~\eqref{running_Renormalon}} \longrightarrow \propto\int \frac{d k_E^2}{(k_E^2)^2}\lambda_i(k_E) \,.\label{running_Renormalon2}
\end{align}
With the analytical and recursive solutions shown in Eq.~\eqref{solution} at hand, one can explicitly show the $n!$-growth. This is formally analogous to the one-field case but with the geometrical series in $t$ replaced by a generic one of the form $\sum a_n t^n$. Then one can build the Borel transform in analogy with section~\ref{onefield}.
In order to find where the multi-variable Borel series diverge, one can employ the direct comparison test with the geometrical series in  Eq.~\eqref{Renormalon_1field}, taking $z_{pole}=2/\beta$ in  Eq.~\eqref{onefieldRenormalon} as a benchmark. Thus, following the logic of Eq.~\eqref{running_Renormalon} or \eqref{running_Renormalon2}, a general procedure can be synthesized from  Eq.~\eqref{solution}  for the $\lambda_i$ and $\forall n$:
\begin{eqnarray}\label{recip}
& \mathcal{F}_{n,i}\equiv \frac{n!}{2^n}\text{Coefficient}(\lambda_i(t),t^n) \\ \nonumber
& \mathcal{F}_{n,i}\mapsto \mathcal{B}_{n,i}(z_1,z_2,...,z_N) \\ \nonumber
& |\mathcal{B}_{n,i}(z_1,z_2,...,z_N)|\leq\mathcal{B}_n^{(1)}(z)|_{z_{pole}=2/\beta}\,,
\end{eqnarray}
where $\mathcal{B}^{(1)}_n$ is defined in Eq.~\eqref{Renormalon_1field} and $\mathcal{B}(\lambda_i)= \sum_n^{\infty} \mathcal{B}_{n,i}$ i.e. it is the Borel series associated with the $\lambda_i$ coupling. The Borel transform is defined through the replacement on $\mathcal{F}_{n,j}$ as  $\lambda_j^k \rightarrow z_j^{k-1}/(k-1)!$ for  $k\geq1$ and $j=1,...,N$. For $k=0$, $\lambda_j^0\rightarrow \delta(z_j)$, where $\delta$ is the Dirac's delta function. The absolute value over $\mathcal{B}_{n,i}(z_1,z_2,...,z_N)$ is a consequence that it is in general a complex function, thus one can test the absolute convergence. Strictly speaking, one should only  require the simple convergence. However, for general complex series there are no theorems to establish the convergence. On the other hand, the absolute convergence is over restrictive and thus automatically guarantees that the perturbative renormalization procedure is consistent inside given regions.

It is convenient to parametrize\footnote{This definition is an agreement  with the notion of complete Reinhardt domains for complex multi-variable series, see
Ref.~\cite{jarnicki2008first} for the mathematical fundament. In Sec.~\ref{toy_model} explicit examples of these domains are shown for a toy model.}
\begin{equation}
(z_1,z_2,...,z_N)\equiv (a_1\times R, a_2\times R,...,a_N\times R)\,,
\end{equation}
with $\{a_1,a_2,...a_N\} \in \mathbb{C}$, $R\in \mathbb{R}$ and positive, then the Abel's lemma may be readily applied, since one recasts the series in the form $\sum_0^{\infty} c_n R^n$, with $c_n \in \mathbb{C}$. Varying $a_i$ within $|a_i|\leq 1$, it is thus possible
to study the domain of convergence in term of $R$ with standard methods. As a result, one identifies the singularities and, in particular, at least one $(z_i)_{pole}\in \mathbb{R}$ and positive. In analogy with the original one-coupling case, such  singularities makes ambiguous the
Laplace transform for $\mathcal{B}_{n,i}(z_1,z_2,...,z_N)$ with $N$ variables:
\begin{equation}
\int dz_1dz_2...dz_Ne^{-(\frac{z_1}{\lambda_1} +\frac{z_2}{\lambda_2}+...+\frac{z_N}{\lambda_N})} \mathcal{B}_{n,i}(z_1,z_2,...,z_N)\,.
\end{equation}
In general the Borel transform is a polynomial of the form:
\begin{align}
&\sum_{n=1}^{\infty} \mathcal{B}_{n,i}(z_1,z_2,...,z_N)=\nonumber \\&\left[\left(\sum_{n=1}^{\infty} C_{i,n}^{k_1,N-1}\right)\prod_{j=1}^{N-1}\delta(z_{j}) + ...\right]+ \left[\left(\sum_{n=1}^{\infty} C_{i,n}^{k_1k_2,N-2}\right) \prod_{j=1}^{N-2}\delta(z_{j}) + ...\right]\nonumber\\
& + \left[\left(\sum_{n=1}^{\infty} C_{i,n}^{k_1k_2...k_m,N-m}\right) \prod_{j=1}^{N-m}\delta(z_{j}) + ...\right] +...  + \sum_{n=1}^{\infty} C_{i,n}\,, \label{generalized_borel}
\end{align}
where $k_1k_2,...,k_m$ symbolize the Borel variables that do not appear in the  $\delta$ products, $N-m$ is the order in the $\delta$ polynomial, $i=1,...,N$  is associated with the coupling $\lambda_i(t)$, $n$ is the order in the expansion in $t$, and finally the coefficients $C_{i,n}^{k_1...,N-m}$  denote the terms of the  power series in the $z_i$. Notice that the Dirac $\delta$ functions makes the generalized  Borel transform in Eq.~\ref{generalized_borel} invertible, as it must be.

In summary, for each coefficient in the above polynomial there is an inequality as in Eq.~\eqref{recip}, and
it shall be manifest below how the direct comparison test reduces to the $n$-root Cauchy-Hadamard formula.
The net result is to locate the divergences of the Borel series, eventually on the real axis, hence identifying the renormalons. At least conceptually, this completes the generalization from one-coupling to the multi-coupling case.

\subsection{Leading renormalon contributions}\label{leading_ren}

The construction in Eqs.~\eqref{recip}-\eqref{generalized_borel} is generic, although it may not be easy to estimate the renormalons singularities in practice.
We define leading renormalons associated with the coupling $\lambda_j$, as the singularity of the Borel transform of the relative series in Eq.~\eqref{RGEs}
for $\lambda_i(t)$, when all other couplings are negligible with respect $\lambda_i$.

Let us re-write the solution of RGEs for a given $\lambda_i(t)$ as
\begin{equation}\label{separation}
\lambda_i(t)=\delta_{ji}\lambda_j(0)+\sum_{n=1}^{\infty}t^n (a_n \lambda_j(0) ^2+A_n )\,,
\end{equation}
where $A$'s contain mixed terms with different $\lambda$s $(\lambda_i^{p_i} \lambda_l^{p_l} \lambda_k^{p_k}...)$.

From Eq.~\eqref{recip} and neglecting $z_l$ for $l\neq j$, one gets
\begin{equation}\label{formalBorel}
\prod_{k\neq j}\delta(z_k)\left(\delta_{ji}+\sum_{n=1}^{\infty} a_n (\frac{z_i}{2})^n\right)\,.
\end{equation}
The Eq.~\eqref{formalBorel} belongs to one of the  $\sum_nC_{i,n}^{k_1,N-1}$ terms in Eq.~\eqref{generalized_borel}.

To see where Eq.~\eqref{formalBorel} diverges, we use Eq.~\eqref{recip}:
\begin{equation}\label{comparison_1field_case}
\frac{1}{2^n} |a_n| (z_j)_{pole}^n= \left(\frac{\beta}{2}z_{pole} \right)^n=1\,,
\end{equation}
where in the last step we have replaced $z_{pole}= 2/\beta$ from~ Eq.~\eqref{onefieldRenormalon}. This leads to
\begin{equation}\label{noA}
(z_j)_{pole}=2/\sqrt[n]{|a_n|}\,.
\end{equation}
This expression is independent from the $A$-terms in Eq.~\eqref{separation}. It is worth noticing that Eq.~\eqref{formalBorel} can
be obtained at any order $n$ from a Taylor expansion up to $o(\beta^n)$ of Eq.~\eqref{generalized_borel}, being $\beta$ any $\beta_i^{pq}$\footnote{The Borel replacement in section~\eqref{recip} reduces of one unit  the power of $z$'s in the terms $a_n z_i^n$, while
it reduces more the powers of $z_i's$ in the terms $A_n$'s. Since the Borel transform  does not act on the power of $\beta$'s,
the net result is that lower powers of $\beta$'s coming from $a_n$ are associated with a given power of $z$'s.}.
As a consequence, Eq.~\eqref{noA} behaves as $1/\beta$, thus neglecting any power of $\beta$ larger than -1. This justifies a posteriori the
name leading renormalons.

Finally, these leading renormalons have to be identified as
\begin{equation}\label{leadingRenormalon}
\lim_{n\rightarrow \infty}2/\sqrt[n]{|a_n|}\,,
\end{equation}
that reproduces the result given in~\cite{tHooft:1977xjm} in the one-field limit, although now it is a more general expression applying to any coupling in a given theory.

Let us explicitly check whether the succession in Eq.~\eqref{formalBorel} is convergent.
It is sufficient to notice that Eq.~\eqref{formalBorel} is an arrangement of the Cauchy's formula, and $a_{n}<a_{n-1}$ due to the content of the small $\beta$s powers\footnote{The limit in   Eq.~\eqref{leadingRenormalon} is non-null unless  $a_n \sim e^{n^{1+\epsilon}}$ with $\epsilon>0$,  but this is not possible since $a_n$ is a subset of Eq.~\eqref{solution} and $a_n<e^n$}.

Bottom line: the concept of leading renormalons provides a generalization of the 't Hooft's result when one coupling can be considered large with respect to the others.

\subsection{Non-leading case and power counting}\label{nonleading_ren}

In the case that one is out of the above hypothesis, the situation becomes less clear. Indeed, there might be cancellations in the Borel series between coefficients of different variables(couplings) whenever they are of the same order. The full analysis of this situation requires a specified framework, i.e.
a model with a given number of fields and couplings. While we shall give such an example below, here we point out that a general insight may still be provided. Merely at the level of power-counting, one can describe the Borel series in terms of the powers of $\beta$s at a given order.
Here we identify with $\beta^n$ any generic power of $\beta_{i j}^n$ (same logic for $z^n$ that now represents any power of $z_i^n$),
and the power counting shows how the contributions to the Borel series grow at order-$k$ in terms of the powers $\beta^n$.
This approach is nothing but the limit in which all the couplings are of the same size.

Consider a generic term of the RGEs that provides the $z^k$ order of Borel series (this generic term belongs to the $C_{i,n}$ term  without Dirac delta functions in the Borel transform shown in Eq.~\eqref{generalized_borel})
\begin{equation}\label{}
\lambda_1^{p_1}...\lambda_N^{p_N},\quad  p_i\geq1\,,
\end{equation}
as usual being $N$ the number of couplings. Thus one has
\begin{equation}\label{}
p_1+...+p_N=k+N\,.
\end{equation}
The corresponding power is $\beta ^{k+N-1}$ and the number of  contributions follows from all the possible combinations producing $k$-order: $\binom{N+k-1}{k}$.
Finally, in agreement with Eq.~\eqref{recip}, the above expression has to be multiplied by $(N+k-1)!/2^k \times z^k/k!$.
Thus, summing up all the possibilities, one gets
\begin{equation}\label{powercounting}
\mathcal{C}(k) \sim \sum _{i=1}^N \frac{(i+k-1)! \beta ^{i+k-1}}{2^{i+k-1} k!} \binom{i+k-1}{k}\,,
\end{equation}
where we denote as $\mathcal{C}(k)$ the generic coefficient of $z^k$. Clearly, one should consider that in some cases $p_i=0$ and when so,
no contribution for the corresponding $i$ appears in the summation in Eq.~\eqref{powercounting}. This would be automatically taken into account by proper coefficients
for any $i$ inside the summation, however such details are beyond the naive power counting argument here presented.
Nevertheless, Eq.~\eqref{powercounting} illustrates how the procedure in Eq.~\eqref{recip} converges.
In fact one has
\begin{equation}\label{ren_k}
\mathcal{C}(k)z_{pole}^k=1 \Rightarrow z_{pole}=1/\sqrt[k]{\mathcal{C}(k)} \,,\,\,\,\,\,\,\, \forall \,k\,.
\end{equation}
Similarly to the leading renormalon discussed above, again we are using implicitly the Cauchy's convergence criterium.
Finally, since
\begin{equation}
\lim_{k\rightarrow \infty} \left[ \frac{(i+k-1)! \beta ^{i+k-1}}{2^{i+k-1} k!} \binom{i+k-1}{k}\right]^{1/k}= \beta/2\,,
\end{equation}
the Eq.~\eqref{ren_k} yields to
\begin{equation}\label{full_ren}
z_{pole} \sim \frac{2}{N \beta}\,.
\end{equation}
This result shows how the number of couplings $N$ affects Eq.~\eqref{onefieldRenormalon} and is valid when the couplings are of the same order. In particular, a large $N$ tends to push the renormalon poles toward smaller values with respect the one-coupling case, therefore worsening the issue of a consistent perturbative renormalizability for smaller coupling constant values. Notice that, since the $z_i$ are assumed to be of the same order and positive (as well as  $\beta$), here the Borel series diverges even in simple sense unlike the general case discussed in Sec.~\ref{multi-field-theory}, in which only absolute convergence may be rigorously proven.

Once again, we stress that Eq.~\eqref{full_ren} is valid at the level of  power-counting.  Therefore,  a dedicated  analysis has to be developed with a specific model at hand, as we shall do in the next section.

\section{A toy model: two coupled-real-scalar field}\label{toy_model}

So far our discussion has been general and to be more specific, we go through a model.
We consider the  minimal extension of the  one-field case, namely a toy model with two real scalar fields $\phi_{1,2}$ and
a global symmetry $Z_2$ (e.g. $\phi_1\rightarrow\phi_1$, $\phi_2\rightarrow -\phi_2$), such that the quartic part of the potential reads as
\begin{equation}
V(\phi_1,\phi_2) \supset \lambda_1 \phi_1^4 +\lambda_2 \phi_2^4 +\alpha \phi_1^2\phi_2^2\,,
\end{equation}
and the general formalism in~\cite{Cheng:1973nv} yields
\begin{align}\label{toy_RGE}
\lambda_1'&= \frac{1}{16\pi^2} ( 72\lambda_1^2 + 2\alpha^2)=\beta_{11}\lambda_1^2 +\beta_{13} \alpha^2\,,  \\ \nonumber
\lambda_2'&= \frac{1}{16\pi^2} ( 72\lambda_2^2 + 2\alpha^2)= \beta_{21}\lambda_2^2 +\beta_{23} \alpha^2\,, \\  \nonumber
\alpha'&=\frac{1}{16\pi^2} [  16\alpha^2 + 24\alpha(\lambda_1+\lambda_2)]=\beta_{33}\alpha^2 +\beta_{31} \alpha\lambda_1+\beta_{32}\alpha\lambda_2\,,
\end{align}
where prime denotes derivative with respect $t\equiv \ln (\mu/\mu_0)$.

There are three different series since there are three couplings.
In the same spirit of the preceding discussion, we deal first with the limit when one of the $z_i$ is much larger than the others, i.e. the leading renormalon. Afterwards, we discuss the general case thereby included the limit with the $z_i$ all of the same size, as in Sec.~\ref{nonleading_ren}).

 \begin{figure}\centering
     \includegraphics[scale=0.6]{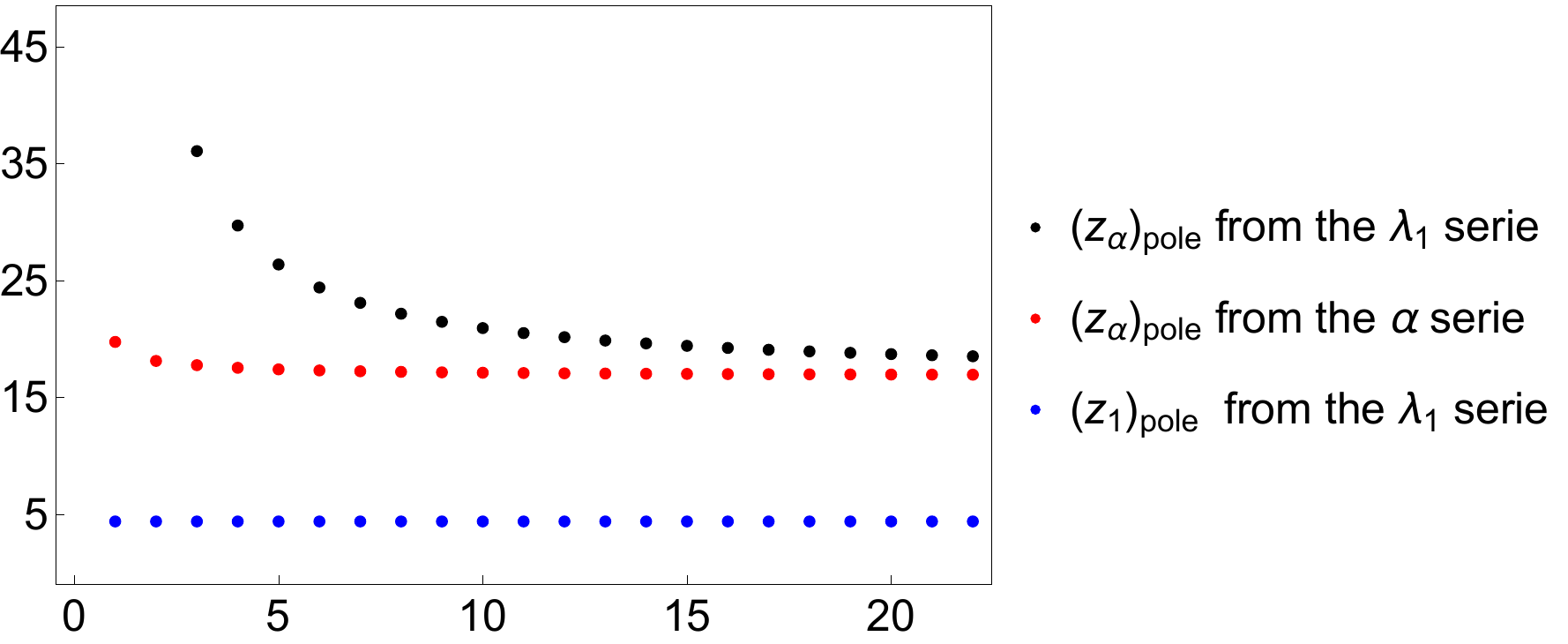}
  \caption{  Asymptotic  value of $(z_1)_{pole}$ and $ (z_{\alpha})_{pole} $ at a given order in $z_i^n$ below  which the Borel series $\mathcal{B}(\alpha)$ and $\mathcal{B}(\lambda_1)$  converge absolutely. The blue dots represents the renormalon for the one field case done in \cite{tHooft:1977xjm}.}
  \label{fig:lead_ren_alpha}
\end{figure}

\paragraph{Leading renormalon case}

Let us denote the solutions given by Eq.~\eqref{solution} as  $\lambda_1(t)$, $\lambda_2(t)$ and $\alpha(t)$.
Following Sec.~\ref{multi-field-theory}, one can implement the Borel transform for these RGEs series and find the expression for their Borel transform up to a given order $t^n$.  In agreement with Eq.~\eqref{recip}, we denote  $\mathcal{B}(\lambda_1)$, $\mathcal{B}(\lambda_2)$ and $\mathcal{B}(\alpha)$ the Borel transforms associated to $\lambda_1(t)$, $\lambda_2(t)$ and $\alpha(t)$ respectively. By applying the  Cauchy's criterion for absolute convergence as in Subsection~\ref{leading_ren}, one finds the asymptotic values for the poles in the Borel space $\{z_1,z_2, z_{\alpha}\} $ to an arbitrary precision. The results are summarized in Fig.~\ref{fig:lead_ren_alpha}.

For the $\mathcal{B}(\lambda_1)$ series, the leading renormalon is given trivially by the well known one-field result~\cite{tHooft:1977xjm} $z_1 < 2 / \beta\approx4.38$ ($\beta= \frac{72}{16\pi^2}$), since in the limit of small $\lambda_2,\alpha$ the Borel series for $\lambda_1$ is just the geometrical series. This is shown in Fig.~\ref{fig:lead_ren_alpha} with the blue dots line. Exactly the same conclusion holds for $\lambda_2$ coupling.

Conversely, for the $\alpha$ coupling, the procedure in Sec.~\ref{leading_ren} turns out to be more useful.
We show the result of applying the Cauchy's criterion to the $\mathcal{B}(\alpha)$ and $\mathcal{B}(\lambda_1)$ series up to  $\mathcal{O}(z_{\alpha}^{25})$. Formally the position of the pole in the Borel axes is $(z_{\alpha})_{pole}=\lim_{n\rightarrow \infty} (1/|c_n^{\alpha}|)^{1/n}$. The corresponding curves (red and black) in Figure~\ref{fig:lead_ren_alpha} reach a plateau and clearly converge to $\sim17$. This corresponds to Eq.~\eqref{leadingRenormalon} for this specific model.
Therefore, when the other couplings are $\ll\alpha$, the perturbative expansion is consistent up to values $\ll(z_{\alpha})_{pole}$.
Further details and some analytical insights are provided in Appendix~\ref{SecApp2}.

\paragraph{General  case }

 \begin{figure}[t]\centering
     \includegraphics[scale=0.45]{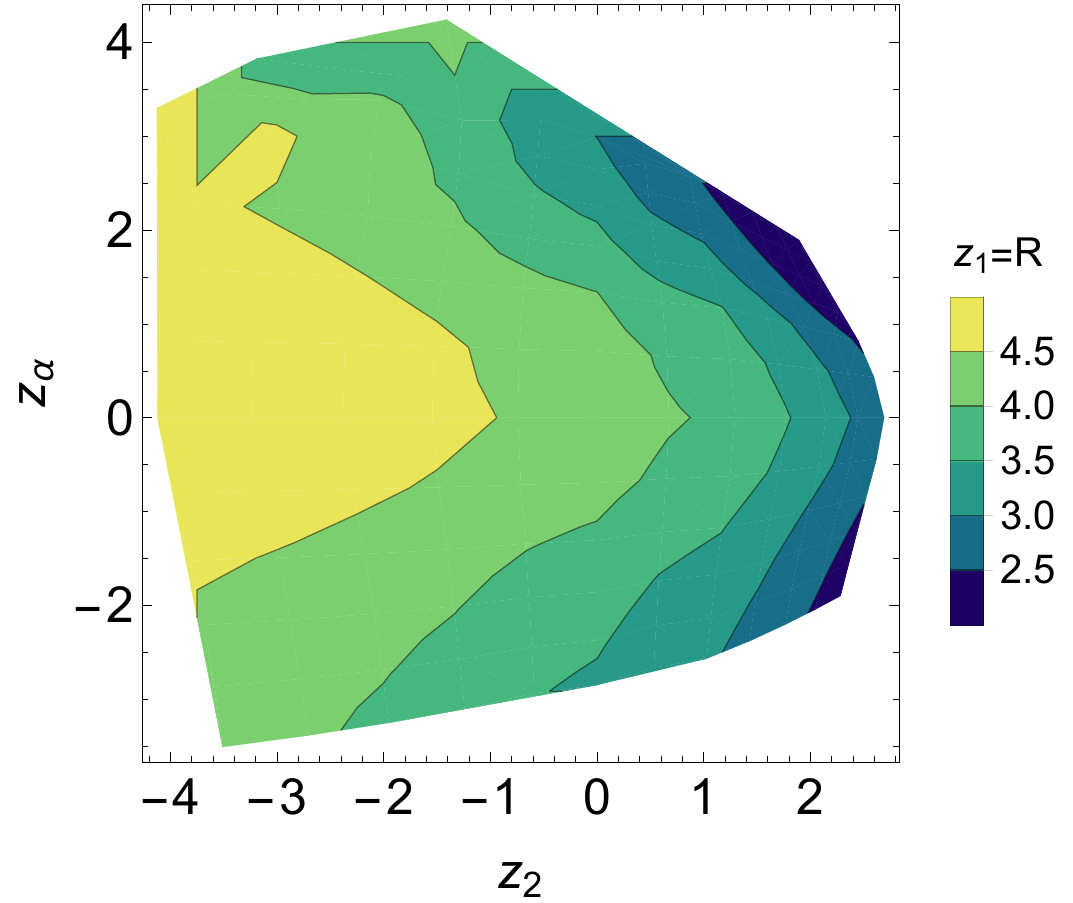}  \hfill   \includegraphics[scale=0.45]{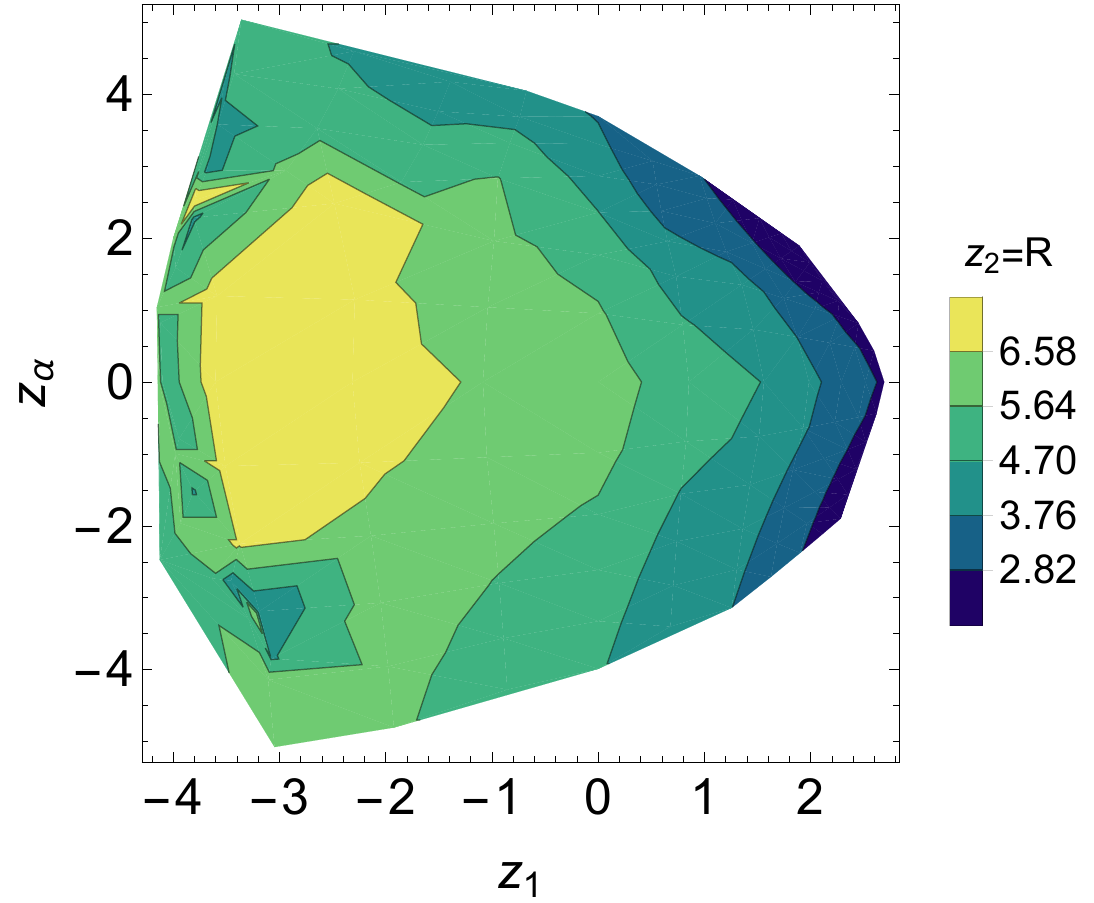} \hfill      \includegraphics[scale=0.45]{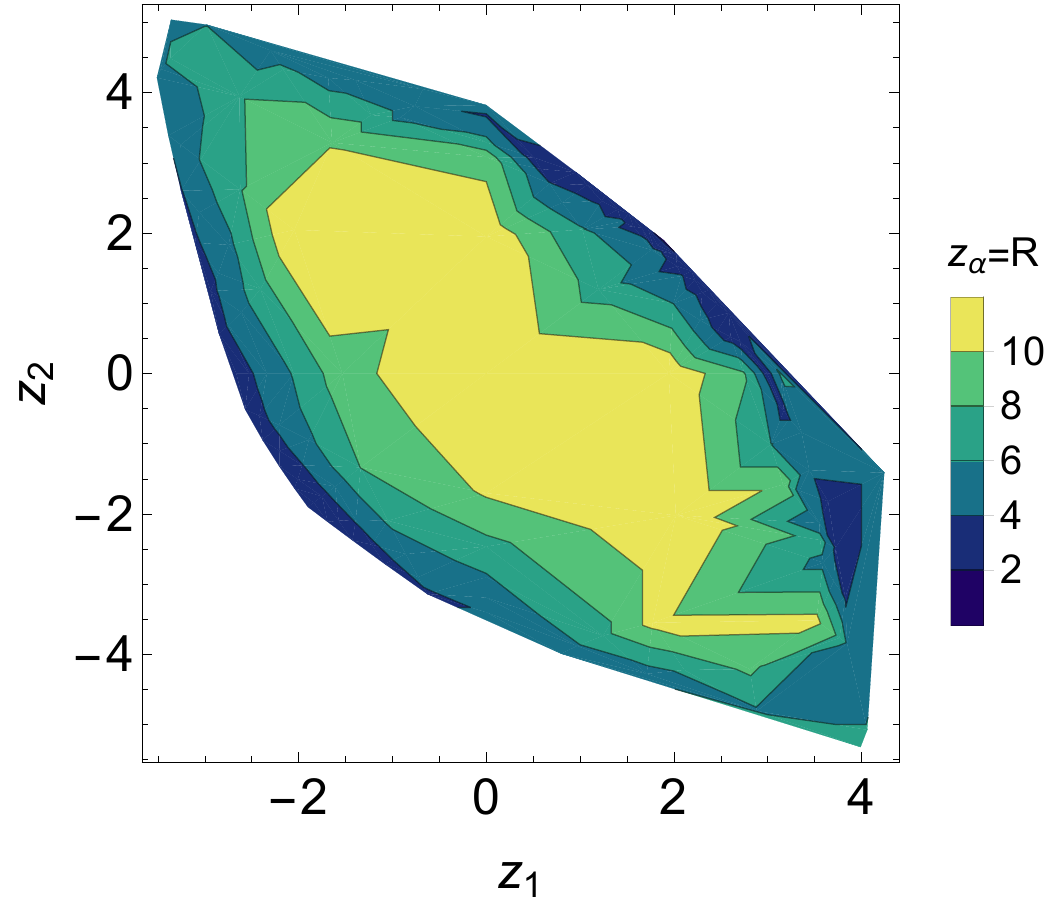}     \\
      \includegraphics[scale=0.45]{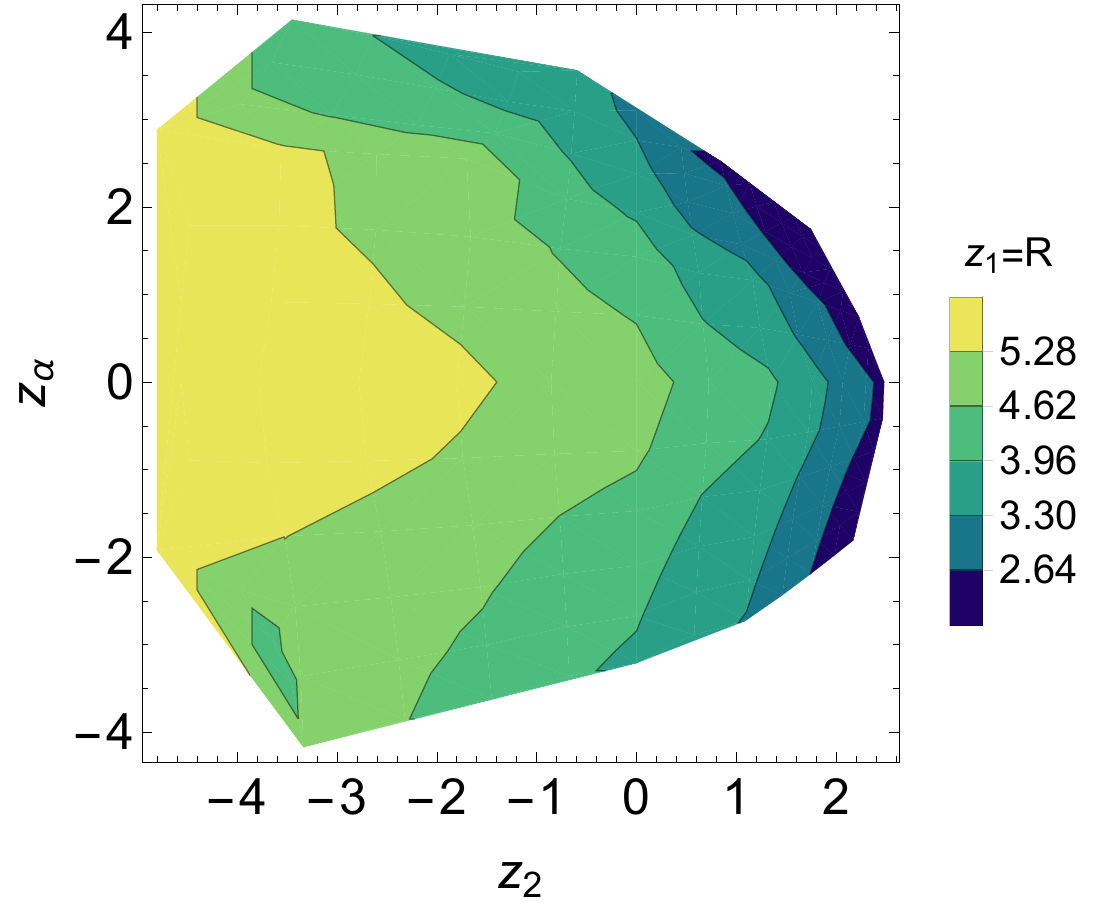} \hfill  \includegraphics[scale=0.45]{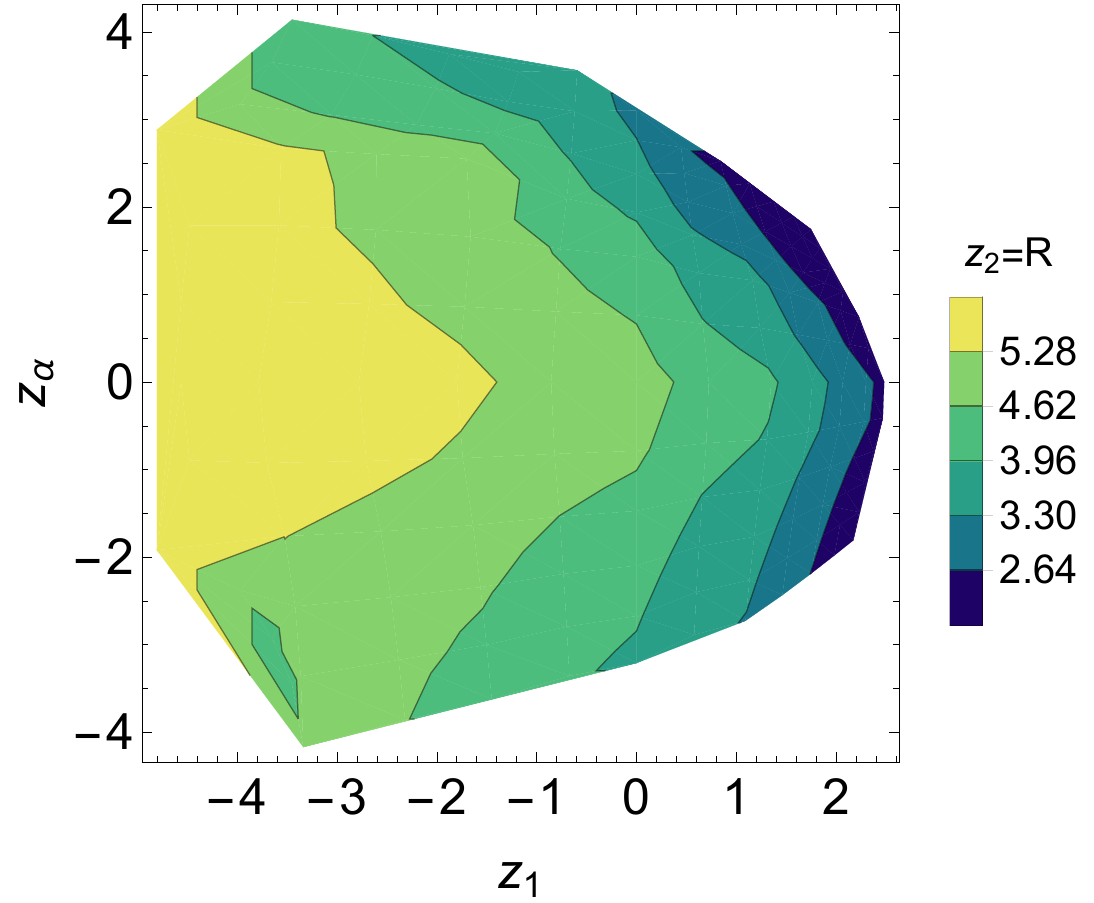} \hfill  \includegraphics[scale=0.45]{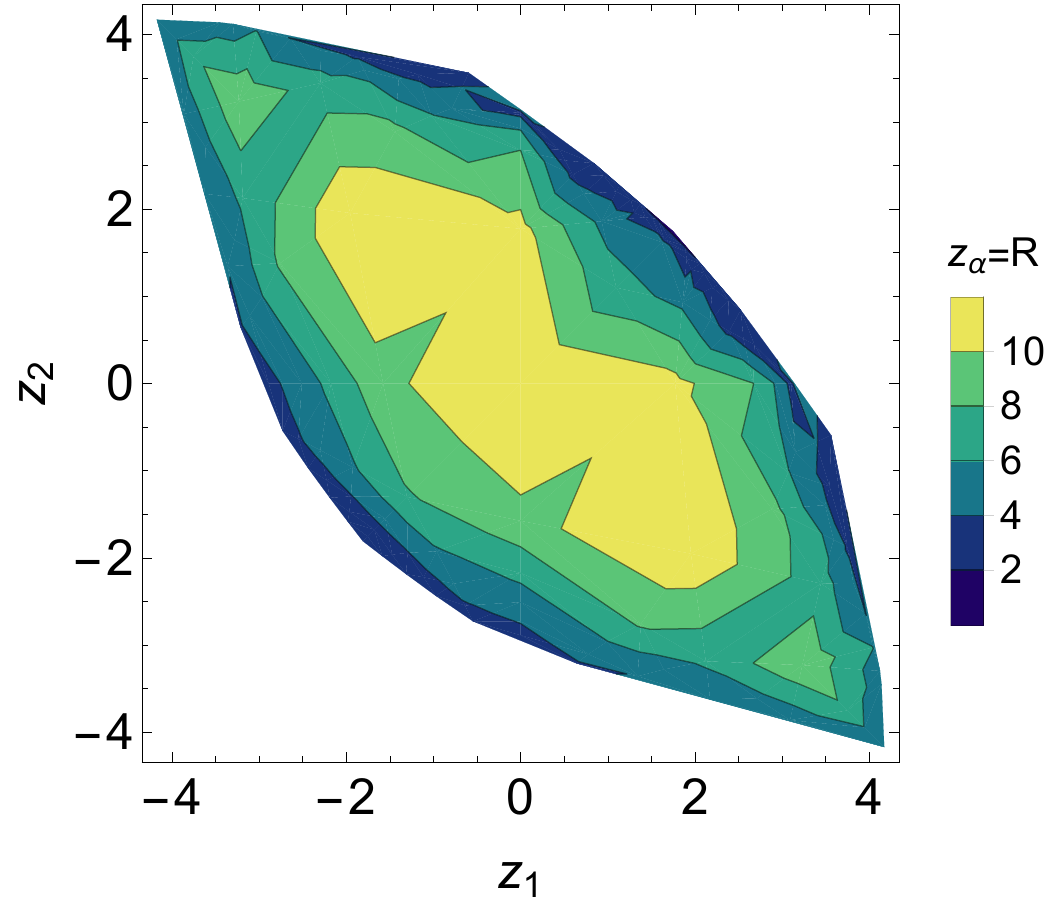}
  \caption{Domains of absolute convergence in the multivariable Real projection space  for the Borel series of the $\lambda_1$ coupling (top) and the $\alpha$ coupling (bottom).
  The regions are convex, as they must for any Reinhardt domain, which is in general proven to be  log-convex~\cite{jarnicki2008first}.}
  \label{fig:orel_z1_bz1}
\end{figure}

Following Section~\ref{multi-field-theory}, we parameterize
\begin{equation}\label{}
z_1=a\times R,\, z_2=b \times R,\, z_{\alpha}=c\times R \,,
\end{equation}
being $z_1$, $z_2$ and $z_{\alpha}$ the Borel variables associated with the couplings $\lambda_1$, $\lambda_2$ and $\alpha$ respectively. The variable
$R \in {\rm I\!R}$ and is positive, the same has to be for at least one of the $a,b,c$ (while the others in general $\in\mathbb{C}$), in agreement with the generalized notion of renormalons in Section~\ref{multi-field-theory}. In particular, we fix the renormalon condition $z_{1,2,\alpha}=R>0,\,(a,b,c=1)$ in turn and identify the respective regions of absolute convergence in $z_i\in {\rm I\!R}$ space (the complex case is discussed in App.~\ref{SecApp2}) for all the Borel series. The result is shown in Fig.~\ref{fig:orel_z1_bz1}. Inside those regions the simple convergence is ensured and hence the perturbation theory  is well defined in the sense of Sec.~\ref{multi-field-theory}. Furthermore, it is clear from Fig.~\ref{fig:orel_z1_bz1} that when  $z_2 \sim z_{\alpha}\sim z_1$, the renormalons are all $ \approx2$ and therefore significantly smaller  than the leading case shown  in Figure~\ref{fig:lead_ren_alpha}. It is worth noticing that this qualitatively agrees with the general power counting argument synthesized in Eq.~\eqref{full_ren}.

\section{Contact with realistic theories}\label{contact}

Our focus on scalar field theories is not only due to illustration purpose. We argue that scalar potentials of realistic theories
might be the central point for issues related to perturbativity and Borel-resummation. While in the Standard Model (SM)~\cite{Weinberg:1967tq,Glashow:1961tr,Salam:1968rm} the main issue is related to the gauge coupling of $SU(3)$, this might no longer be true for models beyond. The Eq.~\eqref{full_ren} shows that the renormalon singularities
are sensitive to the number of couplings and the size of $\beta$s that, in turn, tend to grow depending on the symmetry group and the number of representations.
A common feature of most  beyond-SM (BSM) physics is a non-trivial scalar potential, with a rich field content and a number of different quartics without asymptotic freedom~\cite{thooftasymp,Gross:1973id,Politzer:1973fx}. Whereby the crux of renormalon singularities might be related to scalar quartic couplings too.

Concrete examples can be found already in the simplest extensions of the SM. For instance the doublet-Higgs model(2DHM)~\cite{Lee:1973iz,Inoue:1982pi,Flores:1982pr,Gunion:1984yn} contains five quartic couplings in the potential. A similar situation occurs  for the supersymmetric SM. Another interesting example and popular BSM scenario is the minimal Left-Right symmetric model (LRSM)~\cite{Pati:1974yy,Mohapatra:1974gc,Mohapatra:1974hk,Senjanovic:1975rk,Senjanovic:1978ev}, whose scalar potential contains thirteen quartic couplings.

Thus the issue of renormalons in the Higgs sector might be important in all these examples, or even
sharpened in Grand-Unified models~\cite{Georgi:1974sy,Pati:1974yy}. The same might be argued in theories with a non-fundamendal Higgs particle, such as the little Higgs models (see~\cite{Perelstein:2005ka} and references therein).

It may also occur that a coupling is constrained to be large in a model because
phenomenological requirements. This is as likely as predictive and testable the model is. Such case may be found in low scale LRSM~\cite{Maiezza:2010ic,Bertolini:2014sua} and the 2HDM when the second Higgs is heavy enough~\cite{Gunion:2002zf}. Often the perturbativity bounds are naively inferred by requiring the couplings to be smaller than $4\pi$. This is why the perturbativity issue was analyzed more systematically in low scale LRSM  in Refs.~\cite{Maiezza:2016bzp,Maiezza:2016ybz}, in which, however,  a fuzzy definition of the perturbative regime was used.
On the other hand, the analysis that we have developed in this paper offers a straight consistency check based on the requirement of the Borel resummability of any renormalizable and perturbatively formulated model.

A further possible benefit of Eq.~\eqref{solution} for a realistic model.  For the $\lambda_i$ couplings, the position of the Landau poles can be found by  applying the standard criteria for absolute convergence and are given by
\begin{equation}
\Lambda_{i,Landau}=\mu_0  \text{exp}[\lim_{n\to\infty} |s(n)_i|^{-1/n}]\,,
\end{equation}
which is independent of the initial renormalization scale $\mu_0$. An attempt for an analytical understanding of the Landau poles has been done in Ref.~\cite{Hamada:2015bra} for the SM extended by a scalar field.  A final comment is in order, the analytical solution of the RGE's in Eq.~\eqref{solution} enormously facilitates the computation of the RG-improved effective potential~\cite{Coleman:1973jx}.

\section{Conclusions}\label{outlook}

In this article, we have extended the original concept of ultraviolet renormalon to theories with an arbitrary number of fields and couplings. Our purpose is to find regions in the parameter space of any model where the perturbative renormalisability is guaranteed.
Similarly to the seminal work~\cite{tHooft:1977xjm},
the emergent renormalons can be identified in terms of the $\beta$-function(s). However, the arbitrary number of the fields and couplings sharpen
the issue. In this generic case,  regions in the parameter's space emerge where the perturbative renormalization procedure may fail, because of divergences in the multi-variables Borel
series. Here we have provided a method to find such singularities both analytically (with some assumptions) and numerically, and then to infer bounds on the couplings of a given model.

Furthermore, we have studied how the renormalons behave within a theory, which is usually characterized by a gauge symmetry group and a number of representations.
The renormalons in multi-field theories are affected by both these things: the latter is directly related to the number of couplings $N$;
the former drives both the size of $\beta$'s and $N$. Thus, the renormalons problem tends to be more important when the theory under consideration becomes more involved.

Particular care is reserved to scalar field theory. The practical reason is that we have used a simple toy model with two coupled scalar fields in order to illustrate
the emergence of the renormalons. Our example is the minimal generalization of the original concept proposed in~\cite{tHooft:1977xjm}. Although this toy model is relatively simple,
it is sufficient to test our method and to show how the difficulty of identifying the renormalons increases drastically in the multi-coupling case. The subject of scalar field theory has not only been chosen for the sake of illustration. We argued in fact that in BSM one often deals with an involved scalar sector with many scalar fields and couplings.

In summary, any theory might be perturbatively ill-defined even below the Landau poles. To our knowledge, there is no a  quantitatively precise measure of the non-perturbative regime of a generic QFT. We have tried to fill up this gap by requiring the Borel resumability in a framework with an arbitrary number of fields and couplings. This enables one to determine safe regions in the parameter space of  a given  model, where perturbation theory can be consistently used and resummed.

%
%
\section*{Acknowledgements}

We thank Fabrizio Nesti and Goran Senjanovi\'c for useful comments and for a careful reading of the manuscript.
AM was partially supported by the H2020 CSA Twinning project No. 692194, ''RBI-T-WINNING''. JV was funded by  Fondecyt project N. 3170154 and partially supported by  Conicyt  PIA/Basal FB0821.

%
%
\appendix

\section{ Further details  of RGEs  }\label{SecApp1}

\paragraph{RGEs analytic solution vs. the numerical solution of the differential equations}

We compare the expressions obtained in Sec.~\ref{RGEs} with the numerical solutions of the RGEs equations discussed in Sec.~\ref{toy_model}.

We use for this example the initial condition $\lambda_1(0)=\lambda_2(0)=\alpha(0)=0.1$. The analytical solution in Eq.~\eqref{solution} describes the solution with good accuracy already in the first few orders in the expansion in the variable $t=\log(\mu/\mu_0)$.

\paragraph{General RGEs analytic solution}\label{generalRGE}

Here we discuss the analytic solution of the RGEs in the general case including cubic and quartic terms, since these terms appear when fermions are included into the analysis. Consider scalar fields and write the RGEs at 1-loop in the general form and in the same notation of Sec.~\ref{RGEs}
\begin{align}\label{}
 \lambda_i'= \beta_i^{nm} \lambda_n \lambda_m+\gamma_i^{mns}\lambda_m\lambda_n\lambda_s+\theta_i^{mnsu}\lambda_m\lambda_n\lambda_s\lambda_u\,.
\end{align}

The generalized variational method~\cite{he2000variational} yields
\begin{align}\label{rec_int2}
\lambda_i^{(n)}(t)& = \lambda_i^{(n-1)}(t) - \int_0^t \biggl( \frac{d\lambda_i^{(n-1)}}{ds} -  \sum_{j,k}^N \beta_i^{j k} \lambda_j^{(n-1)}(s) \lambda_k^{(n-1)}(s) +\nonumber \\ & \sum^N_{n,m,l=1}\gamma_i^{mnl}\lambda^{(n-1)}_m(s)\lambda^{(n-1)}_n(s)\lambda^{(n-1)}_l(s)+\nonumber \\&
 \sum^N_{n,m,l,u=1}\theta_i^{mnlu}\lambda^{(n-1)}_m(s)\lambda^{(n-1)}_n(s)\lambda^{(n-1)}_l(s)\lambda^{(n-1)}_u(s)\biggr)ds\,.
\end{align}
This is the complete analytical solution of the RGEs at one-loop and it applies to the  case including fermions and gauge fields. As in  Sec.~\ref{RGEs}, the Eq.~\eqref{rec_int2} can be rearranged as
\begin{align}\label{solution2}
\lambda_i(t)=& \lambda_i(t_0)+ \sum_{n=1}^{\infty}s(n)_i \frac{t^n}{n}, \nonumber \\
s(n)_i=& \biggl[\Theta(n-2) \left(\frac{\partial v_i }{\partial \lambda_k} \right)s(n-1)_k+\Theta(n-3) \frac{1}{2!}\left(\frac{\partial^2 v_i }{\partial \lambda_k\partial \lambda_l} \right) \sum_{m,p}s(m)_ks(p)_l \delta_{m+p,n-1}+\nonumber \\
&  \Theta(n-4) \frac{1}{3!}\left(\frac{\partial^3 v_i }{\partial \lambda_k\partial \lambda_l \partial\lambda_r} \right) \sum_{m,p,w}s(m)_ks(p)_l s(w)_r\delta_{m+p+w,n-1} \nonumber \\ & +\Theta(n-5) \frac{1}{4!}\left(\frac{\partial^4 v_i }{\partial \lambda_k\partial \lambda_l\partial\lambda_r\partial \lambda_z} \right) \sum_{m,p,w,x}s(m)_ks(p)_l s(w)_r s(x)_z \delta_{m+p+w+x,n-1}\biggr]\,,
\end{align}
with $s(1)_i \equiv v_i\,$, $\Theta(n)$ is the Heaviside step function, namely $\Theta(n)=0$  if $n<0$ and $\Theta(n)=1$  if $n\geq0$ and
\begin{equation}
v_i \equiv \beta_i^{nm} \lambda_n(t_0) \lambda_m(t_0)+\gamma_i^{mns}\lambda_m(t_0)\lambda_n(t_0)\lambda_s(t_0)+\theta_i^{mnsu}\lambda_m(t_0)\lambda_n(t_0)\lambda_s(t_0)\lambda_u(t_0)\,.
\end{equation}
As usual, the couplings within this operators are defined at the scale $t=t_0=0$ or $\mu=\mu_0$.
 \begin{figure}\centering
     \includegraphics[scale=0.43]{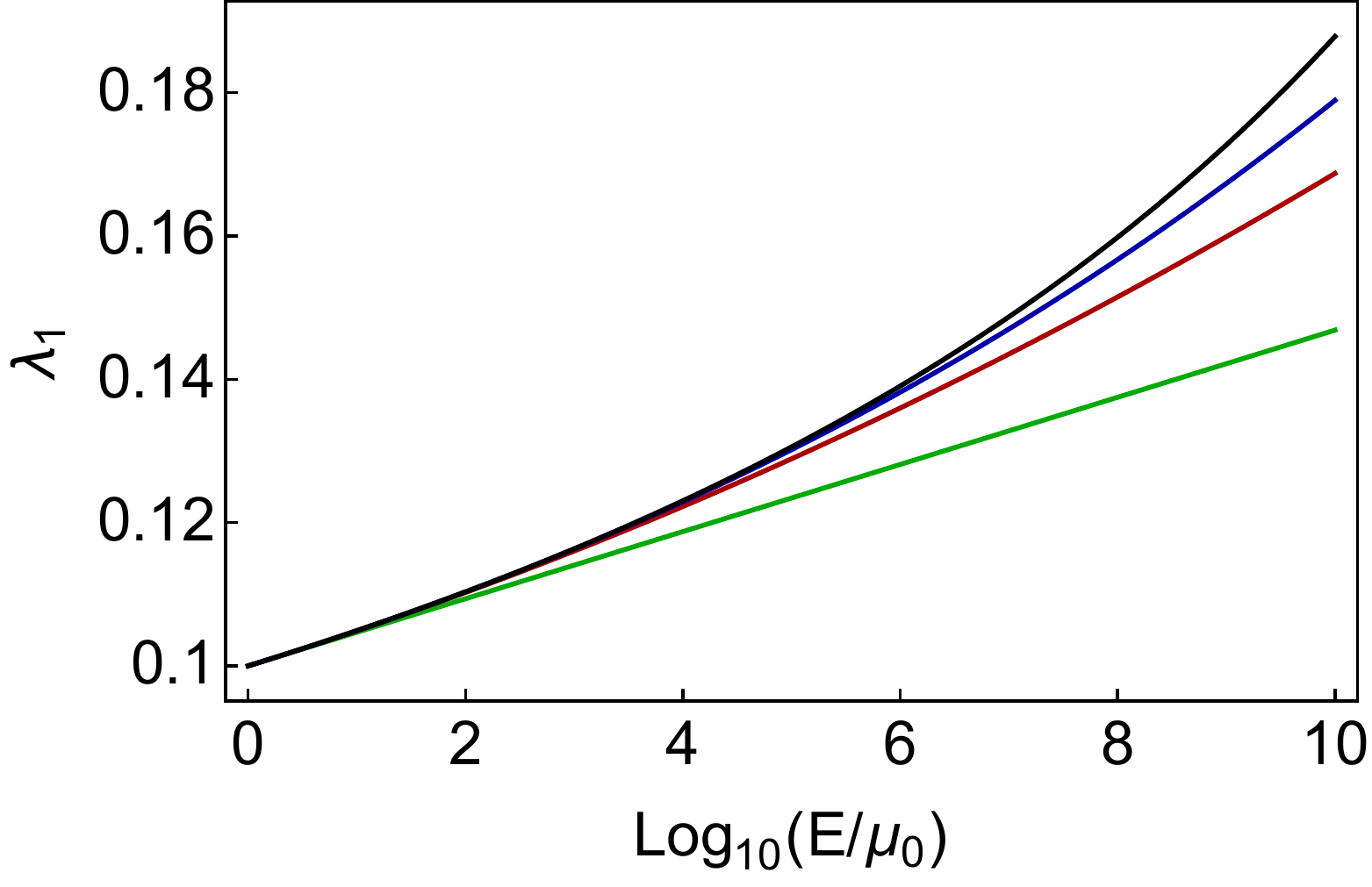}
      \includegraphics[scale=0.43]{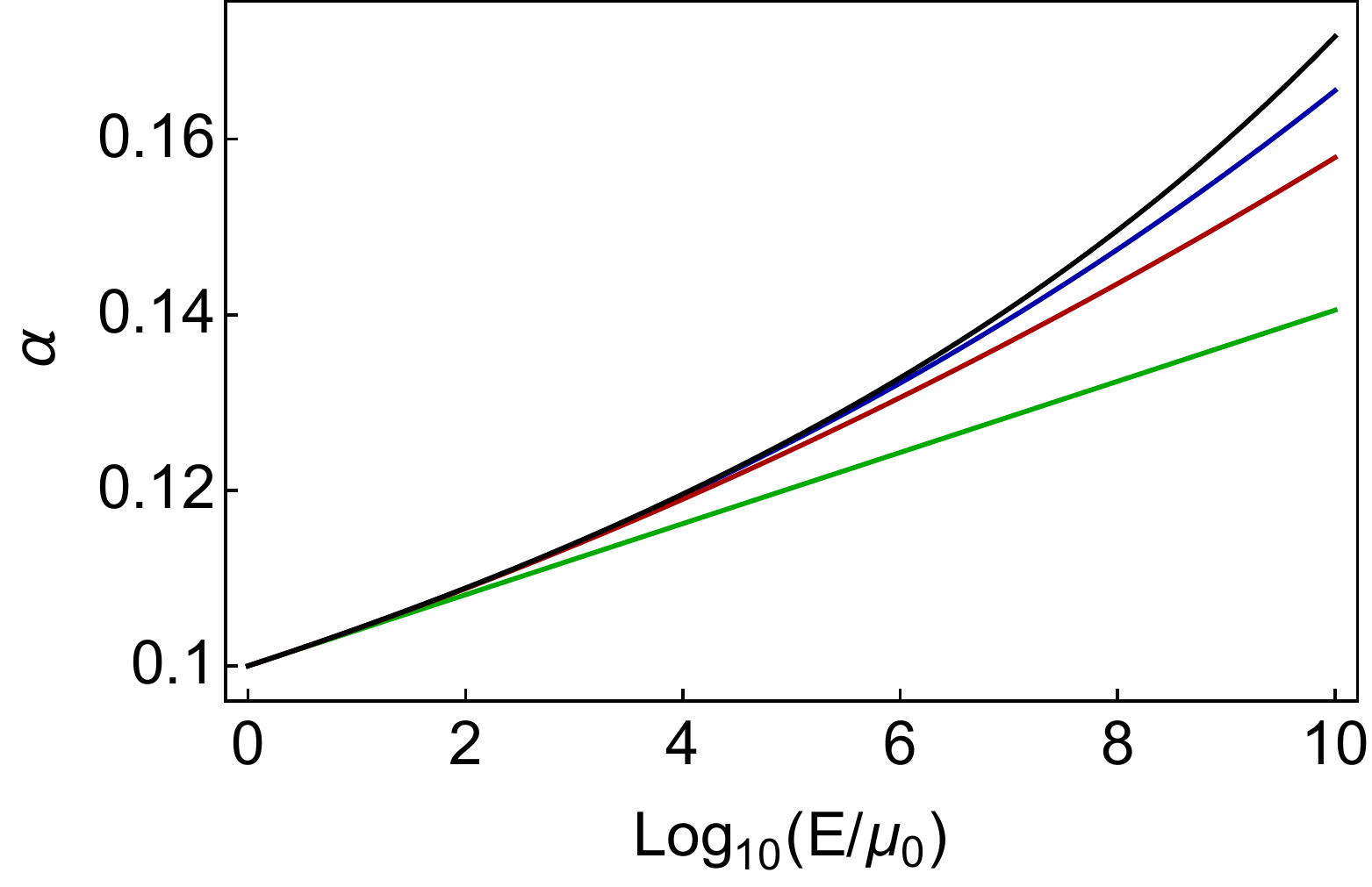}
  \caption{Analytic solution in green, red and  blue at first, second and third order in $t$ versus numerical (black) solution  of the RGEs.}
  \label{fig:comparison}
\end{figure}

\section{Further details of the toy model example} \label{SecApp2}

\paragraph{Analytic insight for renormalons}

We provide some analytical insights of the leading renormalons for the simple model in Sec.~\ref{toy_model}.

As clear from Sec.~\ref{leading_ren}, one can compare  the expression for the Borel series of the coupling $\lambda_1$ with the expression for the one-field case, as in Eq.~\eqref{comparison_1field_case}. This allows us to find the expression of the pole $(z_1)_{pole}$ at any order in $t$. In the specific case of $\lambda_1$ is trivially sufficient the first power of $t$, recovering the well known result of the  one-field(coupling)~\cite{tHooft:1977xjm}
\begin{eqnarray}\label{roughz1}
\delta(z_2)\delta(z_{\alpha})\times (z_1)_{pole}&=&\frac{2}{\beta _{11}}\delta(z_2)\delta(z_{\alpha})\,.
\end{eqnarray}
Same result holds for $\lambda_2$.

The analogous estimation is non-trivial for $\alpha$. An analytical form is still available only for the first low powers in $t$:
\begin{align}
 \delta(z_1)\delta(z_{2})\times(z_{\alpha})^{(1)}_{pole}&= \frac{2}{\beta _{33}}\delta(z_1)\delta(z_{2}) \\
 \delta(z_1)\delta(z_{2})\times(z_{\alpha})^{(2)}_{pole}&= \pm\frac{2\sqrt{2}}{\sqrt{\beta_{13}\beta_{31}+\beta_{23}\beta_{32}+2\beta^2_{33}}} \delta(z_1)\delta(z_{2})\\
 \delta(z_1)\delta(z_{2})\times(z_{\alpha})^{(3)}_{pole}&= \frac{2}{\sqrt[3]{\beta _{33}^3+\beta _{13} \beta _{31} \beta _{33}+\beta _{23} \beta
   _{32} \beta _{33}}}\delta(z_1)\delta(z_{2})\,,
\end{align}
while for high powers in $t$ the numerical result is shown in Fig.~\ref{fig:lead_ren_alpha}. As said in Sec.~\ref{toy_model}, there is no one-field limit for $\alpha$, unlike $\lambda_{1,2}$.

\paragraph{Complex Borel variables}
 \begin{figure}\centering
     \includegraphics[scale=0.45]{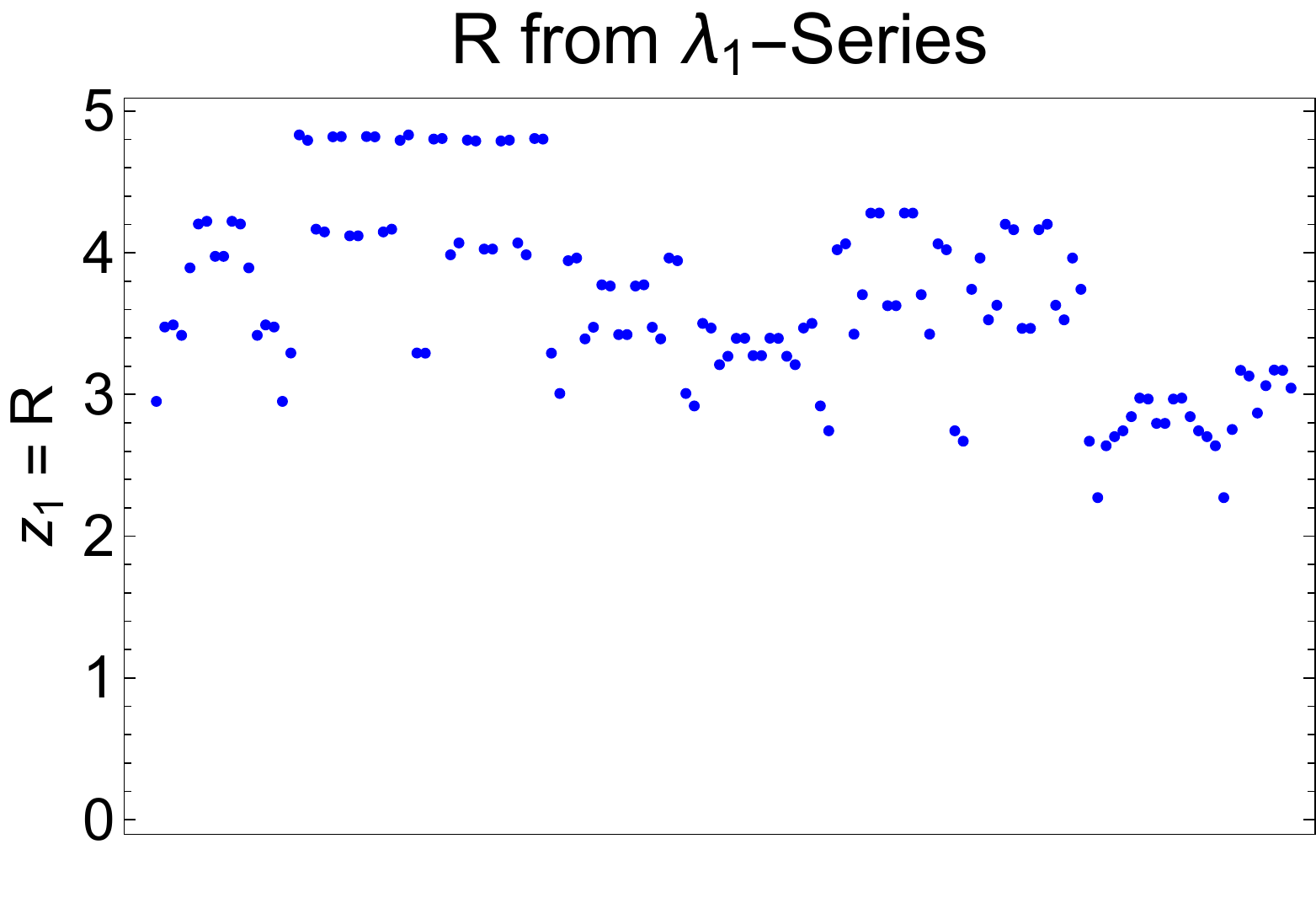}
      \includegraphics[scale=0.45]{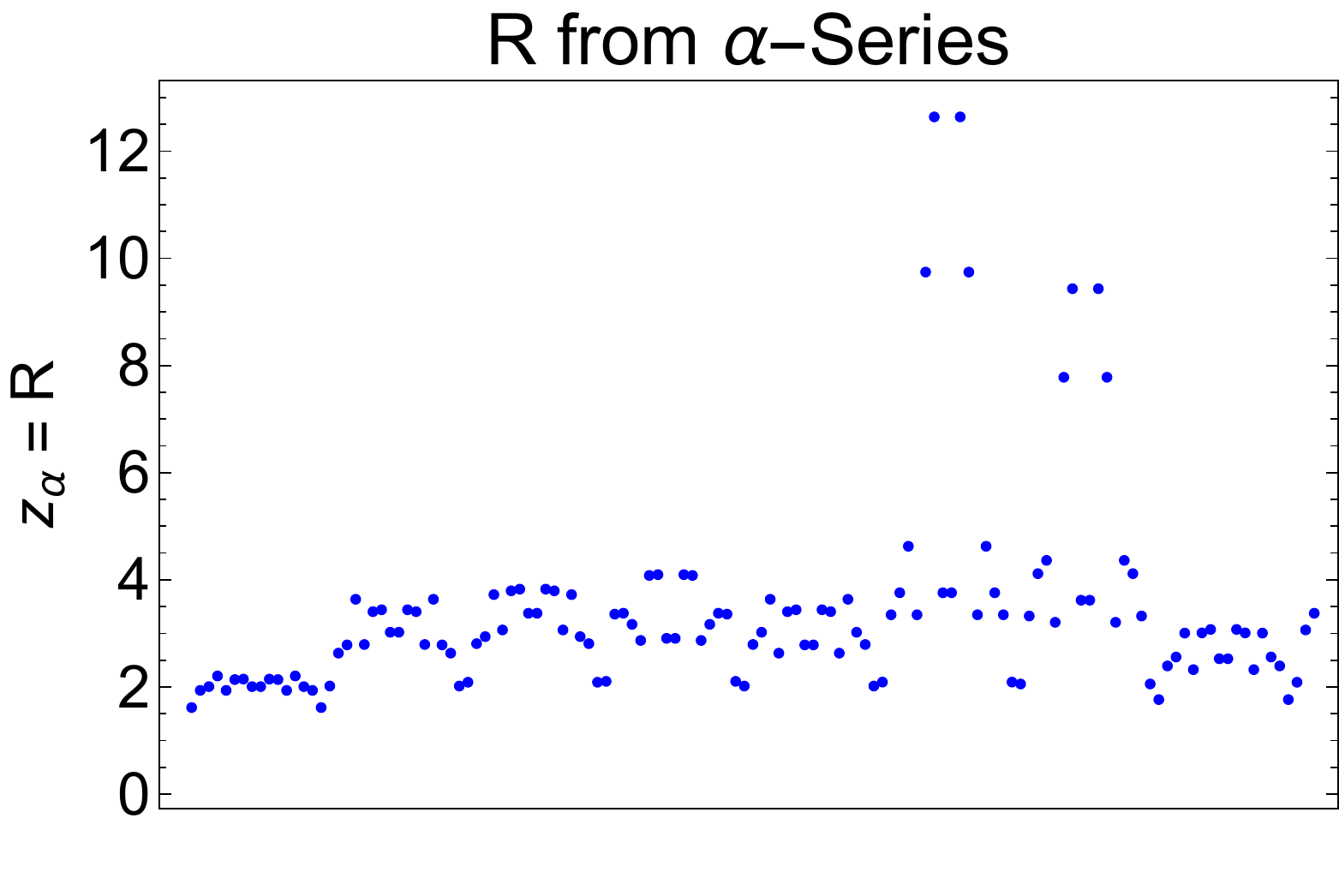}
  \caption{ Left. Scatter plot for the maximum value  $(z_1)_{pole}$ for  complex $z_2$ and $z_{\alpha}$, below which absolute convergence of the Borel series $\mathcal{B}(\lambda_1)$ is obtained. Right.  Scatter plot for the maximum value  $(z_{\alpha})_{pole}$  for complex $z_1$ and $z_2$, below which absolute convergence of the Borel series $\mathcal{B}(\alpha)$ is obtained. }
  \label{fig:orel_z1_bz1_complex}
\end{figure}
Here the case of complex $z_i$ is discussed.
We start with the  $\lambda_1$ series and choose  $z_1\in {\rm I\!R}$, then   $z_2 $ and $z_{\alpha}$ are complex. We parametrize $z_2= b + \tilde{b} i$ and $z_{\alpha}= c + \tilde{c} i$ and calculate the values for $(z_1)_{pole}$ using the  following values for $b,\tilde{b} = \{-1,0.1,0.1,1\}$ and $c,\tilde{c} = \{-1,0.1,0.1,1\}$. There  are  144  combinations and the results obtained for $(z_1)_{pole}$ are shown in Fig.~\ref{fig:orel_z1_bz1_complex} (Left). We see from the plot that
some points for  $(z_1)_{pole}$ are bounded to a maximum value slightly higher than in the leading renormalon case, and that most of the points are smaller than the leading case. It is clear that the renormalon divergence cannot be removed  even when the couplings are complex.
The analysis for $\mathcal{B}(\lambda_2)$ series is identical by construction, due to the symmetry $\lambda_1 \leftrightarrow \lambda_2$ in Eq.~\eqref{toy_RGE}.

In the same way, one can estimate the  value for $(z_{\alpha})_{pole}$ from the $\alpha$ series and therefore we fix $z_{\alpha} = R$ and let $z_1$ and $z_2$ to take complex values. The result is shown in Fig.~\ref{fig:orel_z1_bz1_complex} (Right). Hence the leading renormalon case provides a good estimate on how far the $(z_{\alpha})_{pole}$  can be pushed to the right of the real positive plane.


\bibliographystyle{jhep}
\bibliography{biblio}

\end{document}